# Photon Accelerator in Magnetized Electron-Ion Plasma

S. V. Bulanov[1], S. S. Bulanov[2], T. Z. Esirkepov[3], G. Gregori[4], G. M. Grittani[1],
M. Lamač[1], B. K. Russell[5], A. G. R. Thomas[6], P. Valenta[1]

[1] ELI Beamlines Facility, The Extreme Light Infrastructure ERIC, Za Radnici 835, Dolni Brezany, 25241, Czech Republic
[2] Lawrence Berkeley National Laboratory, Berkeley, California 94720, USA
[3] National Institutes for Quantum and Radiological Science and Technology (QST), Kansai Photon Science Institute, Kyoto, 619-0215 Japan
[4] Department of Physics, University of Oxford, Parks Road, Oxford OX1 3PU, Oxford, UK
[5] Department of Astrophysical Sciences, Princeton University, Princeton, New Jersey 08544, USA
[6] Gerard Mourou Center for Ultrafast Optical Science, University of Michigan, 2200 Bonisteel Boulevard, Ann Arbor, Michigan 48109, USA

## Abstract

Strong magnetic fields and plasmas are intrinsically linked in both terrestrial laboratory experiments and in space phenomena. One of the most profound consequences of that is the change in relationship between the frequency and the wave number of electromagnetic waves propagating in plasma in the presence of such magnetic fields when compared to the case without these fields. Furthermore, magnetic fields alter electromagnetic wave interaction with relativistic plasma waves, resulting in different outcomes for particle and radiation generation. For a relativistic plasma wave-based photon acceleration this leads to an increased frequency gain, and, thus, potentially to higher efficiency. The influence of a magnetic field leads to quantitative and qualitative change in the properties of photon acceleration, amplifying the increase in the electromagnetic wave frequency.

## 1. Introduction

The interaction of intense electromagnetic waves with plasmas of different density and composition almost necessarily leads to the generation of electron density modulations, which can have different properties, including velocities ranging from non-relativistic to relativistic ones. In a rarefied plasma, the relativistic speed of these density modulations is approximately equal to the group velocity of the electromagnetic wave. The interactions of these density modulations with electromagnetic waves, either the ones that generated them or the other ones, lead to changes in the waves' amplitude and frequency. In the case of increasing frequency and amplitude of the wave, this process has been called "photon acceleration", a term introduced by Wilks et al. in [1] 1989 (see also [2], as well as [3-9]). A frequency upshifted and highly intense electromagnetic pulse from a "photon accelerator" and/or an XFEL can be in principle be used to study the otherwise inaccessible regimes of strong instabilities, as well as field quantum electrodynamics [10] in collisions with high-energy electron beams [11].

Typically, when an electromagnetic wave transfers energy into plasma, a process known as wake wave generation occurs. This process belongs to the range of problems associated with the interaction of electromagnetic waves with plasma waves, which, from a more general point of view, includes the study of the transformation of waves of different types and parametric Stimulated Raman scattering of light. Below, we consider only those aspects of photon interactions with wake waves that are directly related to "photon acceleration". Wake waves are actively studied in connection with laser acceleration of electrons within the framework of the laser wake field acceleration (LWFA) paradigm [12-16]. The universal nature of wake acceleration has made it attractive for applications in accelerating high-energy charged particles in space (see [17-19], where this mechanism is considered in relation to cosmic ray acceleration). In fact, the idea of using a wake plasma wave to increase the frequency of electromagnetic waves and amplify them originated in laboratory laser plasmas and was first considered for “photon acceleration” [1] and “relativistic flying mirrors” [20] assuming that electron density modulations in the plasma are driven by a short laser pulse. In what we will consider below, this is a sufficient condition, but not a necessary one, since what is most important is to have a configuration in which the electromagnetic wave interacts with refractive index modulations moving at relativistic speed, which can be in the form of relativistic electron layers, relativistic shock waves, etc. (see [21, 22]). For simplicity, we will use the term "wake wave" in all these cases from now on.

Since the amplification of electromagnetic waves with increasing frequency is studied as one of the main mechanisms of high-energy electromagnetic bursts (for example, in particular, Lorimer radio bursts: [23-25]) generated during supernova explosions and in the magnetospheres of pulsars and magnetars (see [21,22,26-29]), it is advisable to expand the photon acceleration paradigm to construct a theory of the emission of intense coherent radiation during cosmic active processes.

In space environment, plasma is almost always embedded in a strong magnetic field that alters the dispersion equation, which determines the relationship between the electromagnetic wave frequency and wavenumber, compared to the typical case of a relativistic laser plasma [30]. The magnetic field playing important role in relativistic laser plasma [31,32] is known to change wake field acceleration of charged particles [4, 32]. But here we show that it also leads to a quantitative and qualitative change in the photon acceleration process.

## 2. Photon acceleration for linearly polarized electromagnetic waves propagating across a magnetic field

To describe the propagation of a short-wavelength packet of electromagnetic radiation through a plasma, one can use the geometric optics approximation, which is the subject of many works (e.g. see [33-36] and references therein). As well known, geometric optics approximation is closely related to the Wentzel–Kramers–Brillouin (WKB) approximation [37,38]. The equations of the wave-packet motion described by the coordinate $x$ (the center of the wave packet) and momentum $k$ (carrier wave vector) read as follows

$$\dot{k} = -\partial_x \omega \qquad \text{and} \qquad \dot{x} = \partial_k \omega \,. \tag{1}$$

where $\partial_x$ and $\partial_k$ are the partial derivatives with respect to the coordinate and wave number , respectively, and $\omega$ is the wave frequency. In this expression $\partial_k \omega$ is the group velocity $v_g$ of the electromagnetic wave. The wave phase velocity is $v_{ph} = \omega / k$. Here and below the dot stands for the time derivative. The wave packet frequency dependence on the wave number and plasma parameters depends on whether or not the plasma is magnetized.

Below we consider the case when the linearly polarized electromagnetic pulse interacts with collisionless plasma imbedded to the homogeneous magnetic field assuming that the wave vector of the wave and magnetic field are perpendicular to each other. The dispersion equations, which determine the relationship between wavenumber and frequency for linearly polarized electromagnetic waves depend on whether the electric field vector is directed along or perpendicular to the magnetic field.

### *2.1 Ordinary electromagnetic wave*

In the first case of so-called Ordinary Wave (O-wave), when the electric field vector is directed perpendicular to the magnetic field, the dispersion equation is by

$$\omega(x,k,t) = \sqrt{k^2c^2 + \omega_{pe}^2(k_w(x - v_w t))}\,. \tag{2}$$

Here $\omega_{pe}(k_w(x - v_w t)) = \sqrt{4\pi e^2 n(k_w(x - v_w t)) / m_e}$ is the plasma frequency of the wake wave propagating along the $x$-direction with the phase velocity $v_w$, $e$ and $m_e$ are the electron charge and mass, respectively, $n(k_w(x - v_w t))$ is the electron density, and $k_w$ is the inverse scale length of the configuration inhomogeneity. The dispersion equation (2) is the same as in the case of the electromagnetic wave interacting with collisionless unmagnetized plasma [37].

For electromagnetic wave propagating in collisionless plasma there is a known relationship between the group and phase velocities:

$$v_g v_{ph} = c^2\,. \tag{3}$$

#### *2.1.1 Lagrangian of "massive photon". Jacobi integral.*

Using expression (2) we can write equations (1) in the form

$$\dot{k} = -\frac{\omega_{pe}\partial_x\omega_{pe}}{\sqrt{k^2c^2 + \omega_{pe}^2(k_w(x - v_w t))}} \tag{4}$$

and

$$\dot{x} = \frac{kc^2}{\sqrt{k^2c^2 + \omega_{pe}^2(k_w(x - v_w t))}} \tag{5}$$

we can find the wave number and wave frequency as functions of the wave packet velocity $\dot{x}$:

$$k = \frac{(\omega_{pe} / c^2)\dot{x}}{\sqrt{1 - \dot{x}^2 / c^2}} \tag{6}$$

and

$$\omega = \frac{\omega_{pe}}{\sqrt{1 - \dot{x}^2 / c^2}}\,. \tag{7}$$

The equation of the wave-packet motion takes the form

$$\frac{d}{dt}\left[\frac{(\omega_{pe}/c^2)\dot{x}}{\sqrt{1-\dot{x}^2/c^2}}\right] = -\sqrt{1-\dot{x}^2/c^2}\,\partial_x \omega\,, \tag{8}$$

which is the Euler-Lagrange equation

$$\frac{d}{dt}\left(\frac{\partial \mathcal{L}}{\partial \dot{x}}\right) - \frac{\partial \mathcal{L}}{\partial x} = 0\,, \tag{9}$$

for the Lagrangian

$$\mathcal{L} = -\omega_{pe}\sqrt{1-\dot{x}^2/c^2}\,. \tag{10}$$

It has a form of the of the Lagrangian of relativistic particle (e. g. see [39]) having a “mass” equal to the Lagmuir frequency, which depends on the coordinate and time as $\omega_{pe}(k_w(x - v_w t))$.

Below we use dimensionless variables normalized on

$$k_w^{-1}, \qquad (k_w c)^{-1}, \qquad k_w c, \qquad k_w \tag{11}$$

for coordinate $x$, time $t$, frequency $\omega$, wave number $k$, respectively.

Introducing a variable

$$\xi = x - \beta_w t\,, \tag{12}$$

where $\beta_w = v_w / c$ is the normalized phase velocity of the wake wave $v_w$, allows us to obtain the time-independent Lagrangian for photons (i. e. for the short wavelength electromagnetic wave packet)

$$\mathcal{L} = -\omega_{pe}(\xi)\sqrt{1-(\dot{\xi}+\beta_w)^2}\,. \tag{13}$$

Differentiating this Lagrangian with respect to time and using the Euler-Lagrange equation

$$\frac{d}{dt}\left(\frac{\partial \mathcal{L}}{\partial \dot{\xi}}\right) - \frac{\partial \mathcal{L}}{\partial \xi} = 0\,, \tag{14}$$

we obtain

$$\frac{d\mathcal{L}}{dt} = \frac{\partial \mathcal{L}}{\partial t} + \frac{\partial \mathcal{L}}{\partial \xi}\dot{\xi} + \frac{\partial \mathcal{L}}{\partial \dot{\xi}}\ddot{\xi} = \frac{\partial \mathcal{L}}{\partial t} + \frac{d}{dt}\left(\frac{\partial \mathcal{L}}{\partial \dot{\xi}}\right)\dot{\xi} + \frac{\partial \mathcal{L}}{\partial \dot{\xi}}\ddot{\xi} = \frac{\partial \mathcal{L}}{\partial t} + \frac{d}{dt}\left(\frac{\partial \mathcal{L}}{\partial \dot{\xi}}\dot{\xi}\right). \tag{15}$$

In the case of the time independent Lagrangian $\partial \mathcal{L}/\partial t = 0$, this gives the Jacobi integral conservation

$$\frac{d}{dt}\left(\frac{\partial \mathcal{L}}{\partial \dot{\xi}}\dot{\xi} - \mathcal{L}\right) = \frac{d\mathcal{H}_O}{dt} = 0\,. \tag{16}$$

The Jacobi integral is equal to

$$\mathcal{H}_O(\xi,\dot{\xi}) = \frac{\omega_{pe}(\xi)(1-\dot{\xi}\beta_w - \beta_w^2)}{\sqrt{1-(\dot{\xi}+\beta_w)^2}}\,. \tag{17}$$

In terms of the coordinate $\xi$ and momentum

$$\frac{\partial \mathcal{L}}{\partial \dot{\xi}} = \frac{\omega_{pe}^2(\dot{\xi}+\beta_w)}{\sqrt{1-(\dot{\xi}+\beta_w)^2}} = k \tag{18}$$

it is equal to the Hamiltonian

$$\mathcal{H}_O(\xi,k) = \sqrt{k^2+\omega_{pe}^2(\xi)} - \beta_w k\,, \tag{19}$$

which results in the equations of motion

$$\dot{k} = -\partial_\xi \mathcal{H} = -\frac{\omega_{pe}}{\sqrt{k^2+\omega_{pe}^2(\xi)}}\partial_\xi \omega_{pe} \tag{20}$$

and

$$\dot{\xi} = \partial_k \mathcal{H} = \frac{k}{\sqrt{k^2+\omega_{pe}^2(\xi)}} - \beta_w\,. \tag{21}$$

As we noticed above, time independent Hamiltonian is an integral of motion (it is the Jacobi integral),

$$\mathcal{H}_O(\xi,k) = h_O\,, \tag{22}$$

where $h$ is a constant. Using the relationship $k=\sqrt{\omega^2-\omega_{pe}^2(\xi)}$ we rewrite Eq. (22) as

$$\omega - \beta_w\sqrt{\omega^2-\omega_{pe}^2(\xi)} = h_O\,. \tag{23}$$

Its solution can be written in the form

$$\omega = \frac{h_O \pm \beta_w\sqrt{h_O^2-(1-\beta_w^2)\omega_{pe}^2(\xi)}}{1-\beta_w^2}\,. \tag{24}$$

Substituting this expression into Eq. (19) we obtain

$$\dot{\xi} = \sqrt{1-\frac{\omega_{pe}^2(\xi)(1-\beta_w^2)^2}{\left(h_O \pm \beta_w\sqrt{h_O^2-(1-\beta_w^2)\omega_{pe}^2(\xi)}\right)^2}} - \beta_w\,, \tag{25}$$

which allows us to write the solution of the system of equations (20, 21) in quadrature.

#### *2.1.2 Phase plane*

In order to find the maximum frequency upshift, we need to analyze the topology of the phase plane of the dynamical system (20, 21). Contours corresponding to a constant value of the Hamiltonian $\mathcal{H}_O(\xi,k)$ given by Eq. (22) are presented in Fig. 1 a) for a model dependence of the plasma frequency on the coordinate $\xi$

$$\omega_{pe}(\xi) = \omega_{pe}(0)\left(\frac{1+\delta\,\cos(\mathrm{k_w}\xi)}{1+\delta}\right), \tag{26}$$

here $\beta_w = 0.95$, $\omega_{pe}(0) = 1$, $k_w = 1$, and $\delta = 0.5$. (The case, where the dependence of the plasma frequency on the coordinate $\xi$, is given by a nonlinear wake wave, was considered in Ref. [40]). Fig. 1 b) shows iso-contours of the Jacobi integral given by Eq. (23).

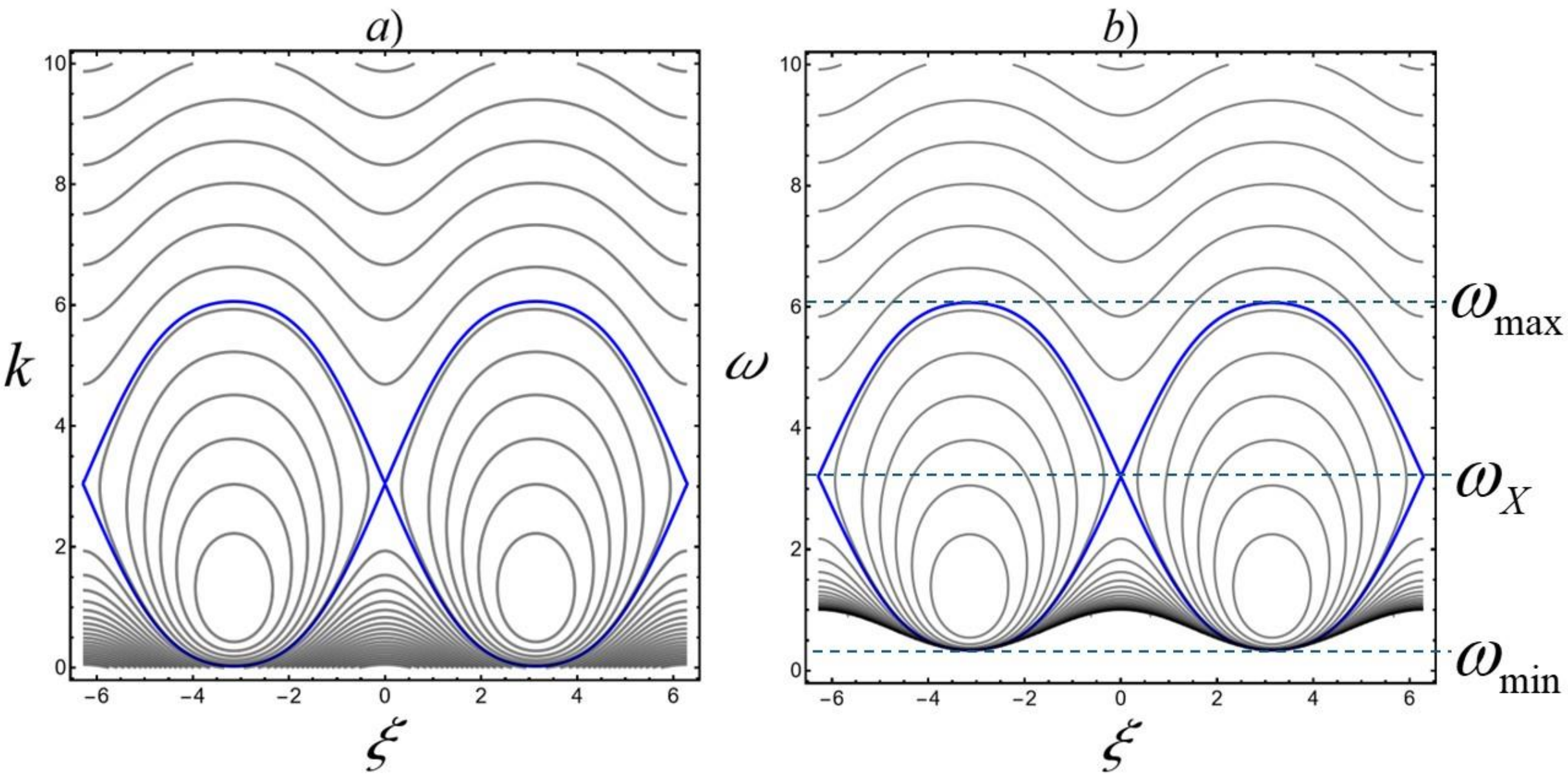

Fig. 1. Contours of constant value of Hamiltonian a) $\mathcal{H}_O(\xi, k)$ given by Eq. (19) and b) of the Jacobi integral $\mathcal{H}_O(\xi, \omega)$ given by Eq. (23) for $\beta_w = 0.95$, $\omega_{pe}(0) = 1$, $k_w = 1$, and $\delta = 0.5$. Here $\omega_{\min}$, $\omega_{\max}$, and $\omega_X$ are the minimal, maximal frequency on the separatrix whose critical point is $(0, \omega_X)$

As one can see, the trajectories of dynamical system described by Eqs. (20, 21) are either trapped or transient in the $(\xi, k)$ plane. The domain of trapped trajectories is encircled by the separatrix whose branches intersect each other in the unstable point of equilibrium,

$$\xi = 2\pi n, \qquad k = \beta_w \frac{\omega_{pe}(2\pi n)}{\sqrt{1-\beta_w^2}}, \tag{27}$$

with $n = 0, \pm 1, \pm 2, \ldots$, where the local group velocity of the electromagnetic wave packet is equal to the phase velocity of the wake wave. The frequency corresponding to the unstable critical point

$$\omega_X = \frac{\omega_{pe}(0)}{\sqrt{1-\beta_w^2}}, \tag{28}$$

is marked in Fig. 1 b) by a horizontal dashed line.

The coordinates of stable points of equilibrium are

$$\xi = \pi(2m+1), \qquad k = \beta_w \frac{\omega_{pe}\pi(2m+1)}{\sqrt{1-\beta_w^2}}, \tag{29}$$

with $m = 0, \pm 1, \pm 2, \ldots$ . On the separatrix the constant on the right hand side (r.h.s.) of Eq. (22) equals

$$h = \omega_{pe}(0)\sqrt{1-\beta_w^2}\,. \tag{30}$$

The phase plane $(\xi, \omega)$ shown in Fig. 1 b) is similar to the $(\xi, k)$ plane with the separatrix determined by the constant in the r.h.s. of Eq. (23) also equal to $h = \omega_{pe}(0)\sqrt{1-\beta_w^2}$ . The coordinates of the critical points on the separatrix in the $(\xi, \omega)$ plane are $\xi = \pi n$ and $\omega = \omega_{pe}(\pi n) / \sqrt{1-\beta_w^2}$ .

The maximum frequency upshift, which is equal to the difference between the maximal and minimal values of the electromagnetic pulse frequency on the trajectory, is reached when the trajectory is the separatrix, i.e. the first integral is equal to $h_O = \omega_{pe}(0)\sqrt{1-\beta_w^2}$ . Substituting it into expression (24) we obtain the maximum and minimum of the electromagnetic pulse frequency on the separatrix

$$\omega_{\max/\min} = \frac{\omega_{pe}(0) \pm \beta_w \sqrt{\omega_{pe}^2(0) - \omega_{pe}^2(\pi)}}{\sqrt{1-\beta_w^2}}, \tag{31}$$

corresponding to plus and minus sign in the r.h.s. of this equation, i. e. the maximum upshifted frequency is

$$\omega_{\max} = \frac{2\beta_w \sqrt{\omega_{pe}^2(0) - \omega_{pe}^2(\pi)}}{\sqrt{1-\beta_w^2}}. \tag{32}$$

For $\beta_w \to 1$ and $\omega_{pe}^2(0) \gg \omega_{pe}^2(\pi/2)$ it is approximately equal to $\omega_{\max} \approx 2\omega_{pe}(0)\gamma_w$, where the gamma-factor equals

$$\gamma_w = 1/\sqrt{1-\beta_w^2}\,, \tag{33}$$

i.e. it formally follows from Eq. (32) it follows that the upshifted frequency tends to infinity at $\beta_w \to 1$. The minimum frequency, which can be considered as a wave frequency before interaction with the wake wave or the frequency of the electromagnetic pulse “injected” in the accelerating phase of the photon accelerator, is

$$\omega_{\min} \approx \omega_{pe}(0)\sqrt{\frac{1-\beta_w}{1+\beta_w}} \approx \frac{\omega_{pe}(0)}{2\gamma_w}, \tag{34}$$

as follows from Eq. (24) for $h_O = \omega_{pe}(0)\sqrt{1-\beta_w^2}$ . If it is expressed in terms of the frequency $\omega_0 = \omega_{\min}$, the maximal upshifted frequency can be written as $\omega_m \approx 4\gamma_w^2\omega_0$ for $\omega_0 = \omega_{\min}$, which is similar to the frequency upshift obtained in the case of a relativistic flying mirror [20, 40]. By virtue of the Lorentz invariance of the ratio of the electric field of the wave to the frequency of the wave, the amplitude of the electromagnetic wave packet also increases proportionally to $4\gamma_w^2$.

In order for a given frequency $\omega_0$ to be approximately equal $\omega_{\min}$, the modulation of the refractive index must be large enough, which leads to the requirement that $\omega_{pe}(0) \geq 2\omega_0\gamma_w^2$ otherwise the trajectory of the electromagnetic wave packet will not be located inside the region bounded by the separatrix. Unlike the case of relativistic flying mirrors, within the framework of the "photon accelerator" concept based on the geometric optics approximation, the frequency shift of an electromagnetic pulse propagating along a trajectory significantly above or below the separatrix wave on a plane is relatively small, since the reflection coefficient of the electromagnetic wave from the wake crest in this case is zero. Another fundamentally important difference between a photon accelerator and a relativistic plasma mirror is that in the case of a photon accelerator, the accelerated electromagnetic pulse propagates in the same direction as the modulation of the refractive index (see Fig. 1b, where the wave number is always positive), whereas in the case of a relativistic mirror, it can collide with the mirror head-on, which provides a large frequency shift (see also the discussion in Refs. [41-43]).

Here, we have analyzed the properties of a photon accelerator in 1D assuming homogeneous unmagnetized plasma. These two assumptions significantly simplify the analysis of the problem, making it possible to provide a clear analytical characterization. In general case, the photon acceleration is a 3D process happening in inhomogeneous plasma. However, the analysis of this process is beyond the scope of the present work. We note that in the case of relativistic mirrors there are several papers on the multi-dimensional analysis of these phenomena taking place in different plasma environments. For example, 3D effects significantly modify the properties of electromagnetic wave interactions with electron density modulations. The focusing of a laser pulse by a paraboloidal wake wave [44, 45] leads to a scaling of upshifed frequency with group velocity different from the 1D case [46-49]. Other spatial and temporal shapes of electron density modulations can be considered to optimize the reflection and intensification efficiencies as shown in Refs. [20, 40, 41, 46]. The plasma density inhomogeneity, on the other hand, can result in the dephasingless regime of the photon acceleration (see Refs. [7-9,50]). The photon accelerator can be also driven by the electron/proton beams [48].

We note that the symmetry of the interaction of an electromagnetic wave and a plasma wave, when dependence of the refraction index on time and coordinate has the form given by Eq. (12), $\xi = x - \beta_w t$, leads to the conservation of the Jacobi integral

$$\omega - \beta_w k = h\,, \tag{35}$$

where $\omega$ and $k$ are the frequency and wave number of the electromagnetic wave packet.

*2.2 Extraordinary electromagnetic wave*

For the Extraordinary wave (it is also called the X-wave) the relationship between the wave frequency and wave number has a form number has a form (e.g. see [51])

$$\frac{k^2}{\omega^2} = 1 - \frac{\omega_{pe}^2(\omega^2 - \omega_{pe}^2)}{\omega^2(\omega^2 - \omega_{pe}^2 - \omega_{Be}^2)}\,. \tag{36}$$

If the wave frequency becomes equal to

$$\omega = \omega_{UH} = \sqrt{\omega_{pe}^2 + \omega_{Be}^2} \tag{37}$$

the denominator in the r.h.s. of Eq. (36) tends to zero, which corresponds to the Upper Hybrid (UH) resonance. The ion component is assumed to be at rest. The electromagnetic wave can propagate provided that the square of the wave number is positive, i.e. in frequency ranges

$$\omega_1 < \omega < \omega_{UH} \qquad \text{and} \qquad \omega_2 < \omega , \tag{38}$$

where

$$\omega_{1,2} = \sqrt{\frac{\omega_{pe}^2 + \omega_{UH}^2}{2} \pm \sqrt{\frac{(\omega_{pe}^2 + \omega_{UH}^2)^2}{4} - \omega_{pe}^4}} . \tag{39}$$

Equations of motion of a short electromagnetic wave (X-wave) packet within the framework of geometric optics approximation have the Jacobi integral

$$\mathcal{H}_{\mathrm{X}}(\xi,\omega) = \omega - \beta_w \sqrt{\omega^2 - \omega_{pe}^2(\xi)\frac{\omega^2 - \omega_{pe}^2(\xi)}{\omega^2 - \omega_{UH}^2(\xi)}} = h_X , \tag{40}$$

For the Jacobi integral $\mathcal{H}_X(\xi,\omega)$ of the X-wave given by this equation the phase plot is presented in Fig. 2. The stripe around the upper-hybrid frequency $\omega_{UH}$, i.e. in the region of the upper-hybrid resonance, subdivides the phase plane in two subdomains. In the case where the Langmuir frequency is approximately equal to the Larmor frequency, $\omega_{pe} \approx \omega_{Be}$, the phase plane is shown in Fig. 2 a). Here the parameters are as follows: the normalized phase velocity of the refraction index modulations is $\beta_w = 0.999$, the Langmuir frequency at the maximum is $\omega_{pe}(0) = 1$, the Larmor frequency equals $\omega_{Be} = 1.25$, and the amplitude of the modulations of refraction index is determined by Eq. (24) with the refraction index modulation amplitude determined by $\delta$. As in Fig. 1, $\omega_{\min}$, $\omega_{\max}$, and $\omega_X$ are the minimum, maximum frequency on the separatrix whose critical point is $(0,\omega_X)$.

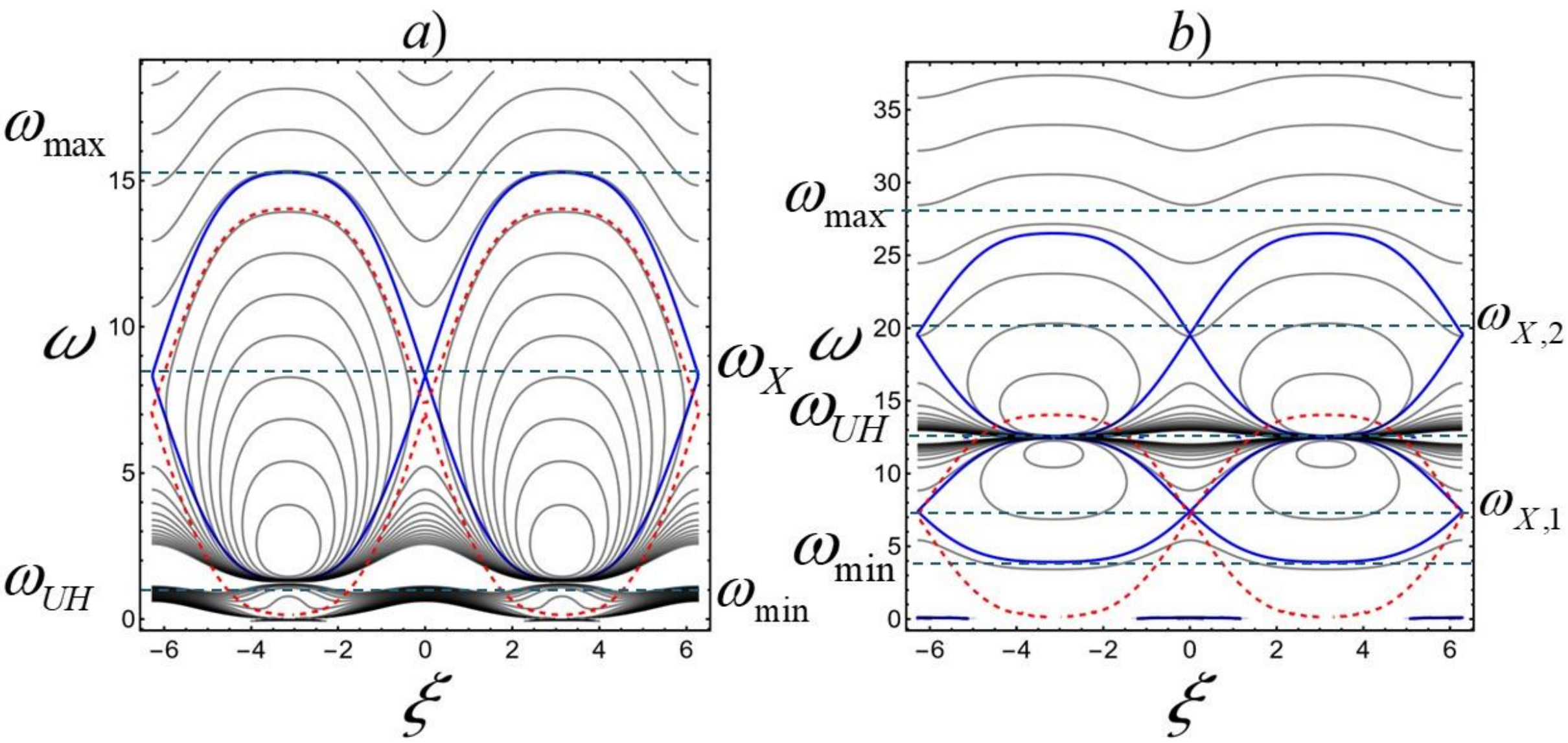

Fig. 2. Contours of constant value of the integral $\mathcal{H}_X(\xi,\omega)$ given by Eq. (40) for the X-wave. Here $\beta_w = 0.999$, $\omega_{pe}(0) = 1$, and the refraction index modulation amplitude is determined by $\delta = 0.5$ and Larmor frequency equal to $\omega_{Be} = 1.25$ in frame (a) and $\omega_{Be} = 12.5$ in frame (b). As in Fig. 1, $\omega_{\min}$, $\omega_{\max}$, and $\omega_X$ and $\omega_{X,1}$ and $\omega_{X,2}$ are the minimal, maximal frequency on the separatrix (blue curve) whose critical point is $(0,\omega_X)$. Red dashed curves show the separatrices for $\omega_{Be} = 0$.

In Fig. 2, the contours of constant value of the Jacobi integral $\mathcal{H}_X(\xi,\omega)$ given by Eq. (40) are presented for $\beta_w = 0.999$, $\omega_{pe}(0) = 1$, where the refraction index modulation amplitude is determined by $\delta = 0.5$ and Larmor frequency equal to $\omega_{Be} = 1.25$ in frame (a) and $\omega_{Be} = 12.5$ in frame (b). Here, as in Fig. 1, $\omega_{\min}$ and $\omega_{\max}$ are the minimal and maximal frequency on the separatrix (blue curve) whose critical point is $(0,\omega_X)$. Red dashed curves show the separatrices for $\omega_{Be} = 0$. As shown in Fig. 2a and 2b, a strong magnetic field $(\omega_{Be} \gg \omega_{pe})$ ensures effective interaction of the electromagnetic pulse with the wake wave, enabling a significantly higher maximum upshifted frequency compared to the case of weakly magnetized plasma.

Within a narrow stripe in the vicinity of the upper hybrid resonance region the X-wave decays via generation of the electrostatic Bernstein wave and subsequent electron heating. We leave this regime discussion to forthcoming papers.

**3. Photon acceleration for circularly polarized electromagnetic waves propagating along a magnetic field**

In this section we consider circularly polarized electromagnetic waves propagating in plasmas in the presence of a strong magnetic field with the goal to demonstrate the difference between the L- and R-wave polarizations (left- or right-rotated electric field). The R-wave corresponds to right-handed circular polarization with the electric field rotating in the same sense as electron gyration. In the L-wave the electric field rotates in the opposite direction. The dispersion equation gives the relationship between the wave number and wave frequency for circularly polarized electromagnetic waves propagating along the magnetic field $B$ (see [51]):

$$\frac{k^2}{\omega^2} = 1 - \frac{\omega_{pe}^2}{\omega(\omega \pm \omega_{Be})}, \tag{41}$$

where $\omega_{Be} = eB / m_e c$ is the electron Larmor frequency. In the normalized form it can be written as

$$\omega_{Be} = \frac{eB}{k_w m_e c^2}, \tag{42}$$

The ion component is assumed to be at rest. The plus and minus signs in the r.h.s. correspond to the left circularly polarized wave (or an L-wave) and to the right circularly polarized wave (or an R-wave) respectively. The latter is often referred to as the whistler mode. The electromagnetic wave can propagate provided that the square of the wave number is positive, i.e., its frequency is bound by the following limits

$$0 < \omega < \omega_{Be} \quad \text{and} \qquad \omega > \frac{1}{2}\ \sqrt{\omega_{Be}^2 + 4\omega_{pe}^2(\xi)} + \omega_{Be} \tag{43}$$

for the R-wave, and

$$\omega > \frac{1}{2}\ \sqrt{\omega_{Be}^2 + 4\omega_{pe}^2(\xi)} - \omega_{Be} \tag{44}$$

for the L-wave. As we see the denominator in r.h.s. of Eq. (41) vanishes for R-wave when the wave frequency becomes equal to the Larmor frequency. This corresponds to the cyclotron resonance between the electron Larmor rotation and the electromagnetic wave. In Fig. 3 we present the square of the refraction index $N^2 = k^2 / \omega^2$ for R- and L waves (black and blue curves) given by the Eq. (41) as a function of the wave frequency $\omega$ for $\omega_{pe} = 1$ and $\omega_{Be} = 1.5.$ As it was mentioned above the condition for the electromagnetic wave to be able to propagate in plasmas is that the square of the wave number is positive or $N^2 > 0$.

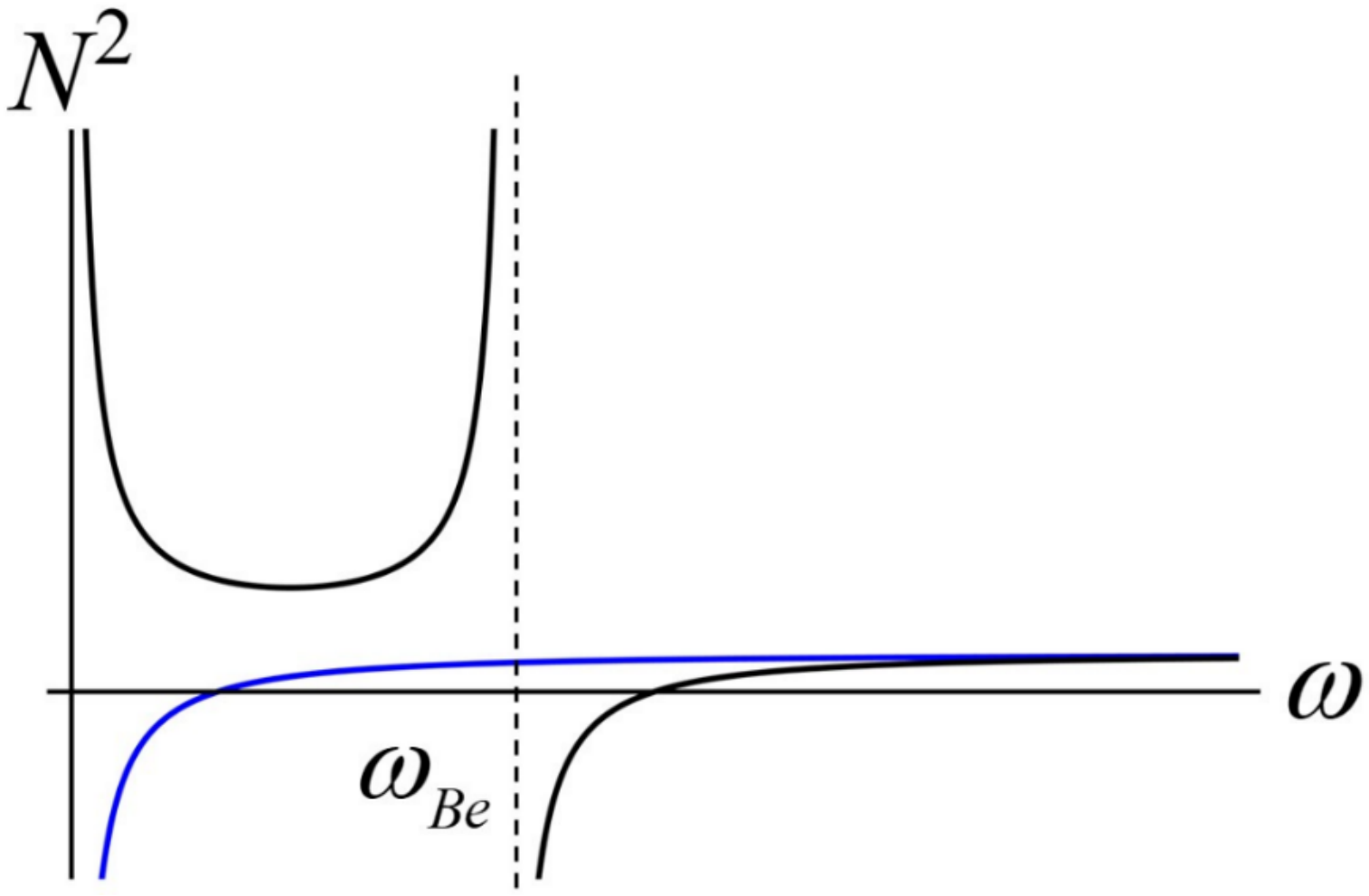

Fig. 3. Square of refraction index for R- and L waves (black and blue curves) given by Eq. (41) as a function of the wave frequency $\omega$ for $\omega_{pe}(0)=1$ and $\omega_{Be}=1.5.$ The waves can propagate in the region of parameters where $N^2>0$.

From the dispersion equation (41) we can find the dependence of the wave frequency on the wave number. However, since the equation (41) is cubic algebraic equation of the third order, it is cumbersome and it is therefore convenient to analyze the $(\xi,\omega)$ phase space using the Jacobi integral $\mathcal{H}_R(\xi,\omega)=h_R$ written in the form

$$\mathcal{H}_R(\xi,\omega)=\omega-\beta_w\sqrt{\omega^2-\frac{\omega\,\omega_{pe}^2(\xi)}{\omega-\omega_{Be}}}=h_R \tag{45}$$

for the R-wave, and

$$\mathcal{H}_L(\xi,\omega)=\omega-\beta_w\sqrt{\omega^2-\frac{\omega\,\omega_{pe}^2(\xi)}{\omega+\omega_{Be}}}=h_L \tag{46}$$

for the L-wave, respectively.

Using dispersion equation (41) we can derive the expression for the group velocity of the R- and L-waves, $v_g=\partial_k\omega$. It is

$$v_g(\omega;\xi)=\frac{2(\omega\pm\omega_{Be})^2\sqrt{\omega^2-\omega\omega_{pe}^2(\xi)/(\omega\pm\omega_{Be})}}{2\omega(\omega\pm\omega_{Be})^2\mp\omega_{Be}\omega_{pe}^2(\xi)}, \tag{47}$$

where the plus sign corresponds to the L-wave and the minus sign corresponds to the R-wave, respectively. The group velocity vanishes at the points given by Eqs. (43) and (44) (see Fig. 4, where group velocities defined by Eq. (47) are shown for $\omega_{pe}(0)=1$). We note that for circularly polarized electromagnetic waves propagating in plasmas along the magnetic field there is no such relationship between the phase and group velocities as in the case of an unmagnetized plasma, which is given by Eq. (3).

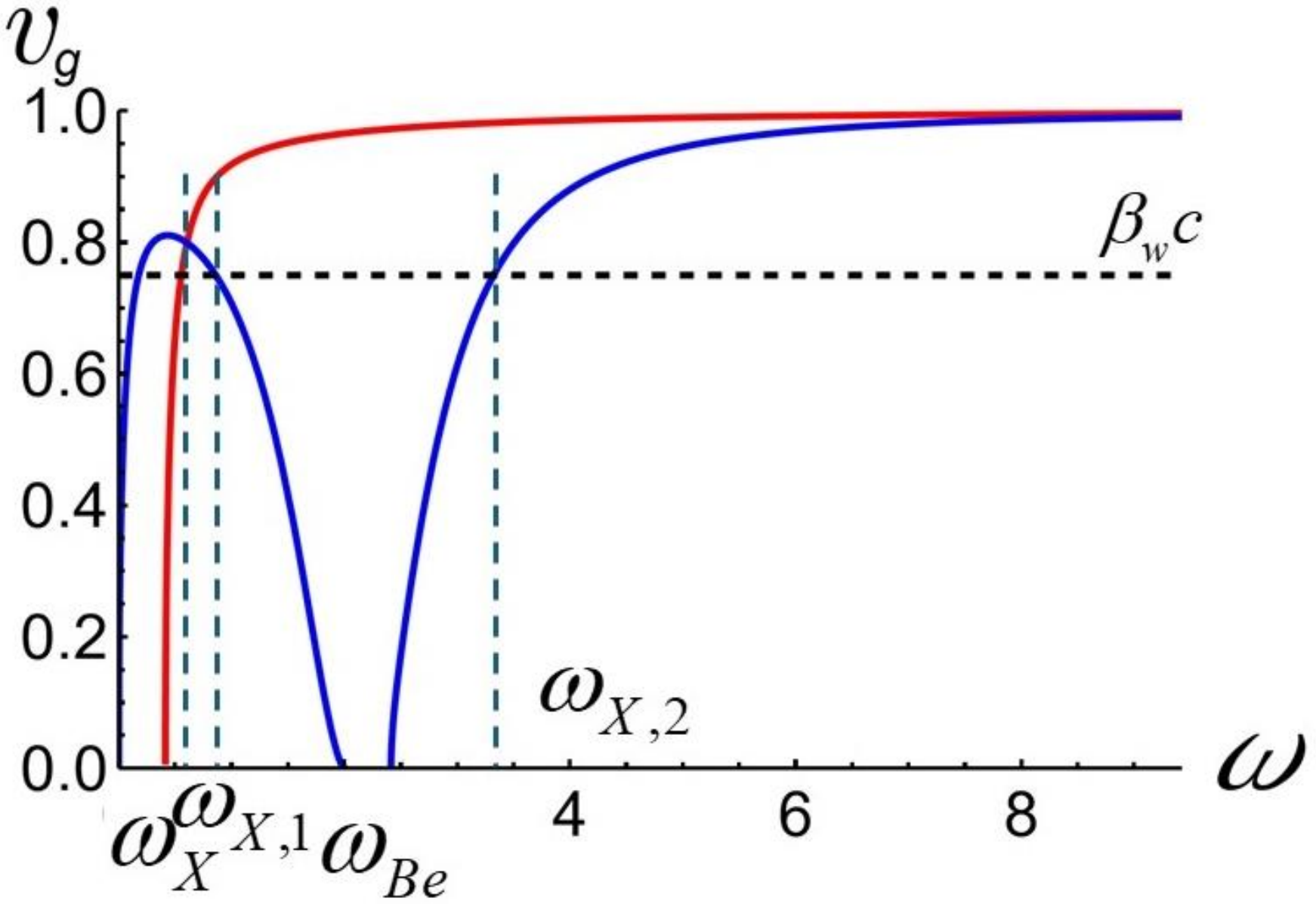

Fig. 4. Group velocity of electromagnetic pulse (red color curve for L-wave and blue color curve for R-wave) vs the wave frequency at $\omega_{pe} = \omega_{pe}(0)$. The points, where the group velocity is equal to the phase velocity of the refraction index modulation, $\beta_w c$ correspond to the critical points on the separatrix $\omega_X$, or $\omega_{X1}$, and $\omega_{X2}$ , where two separatrix branches intersect each other. Here $\beta_w = 0.75$, $\omega_{pe}(0) = 1$, and $\omega_{Be} = 2$.

### *3.1 R-wave*

Contours of constant value of the integral $\mathcal{H}_{\mathcal{R}}(\xi, \omega)$ given by Eq. (45) for the R-wave, in the case when Langmuir frequency is approximately equal to Larmor frequency, $\omega_{pe} \approx \omega_{Be}$, are presented in Fig. 5 a) $\beta_w = 0.999$, $\omega_{pe} = 1$, $\omega_{Be} = 1.25$, and $\delta = 0.75$ . Fig. 5 b) shows a zoom of the region near $\omega = \omega_{Be}$ .

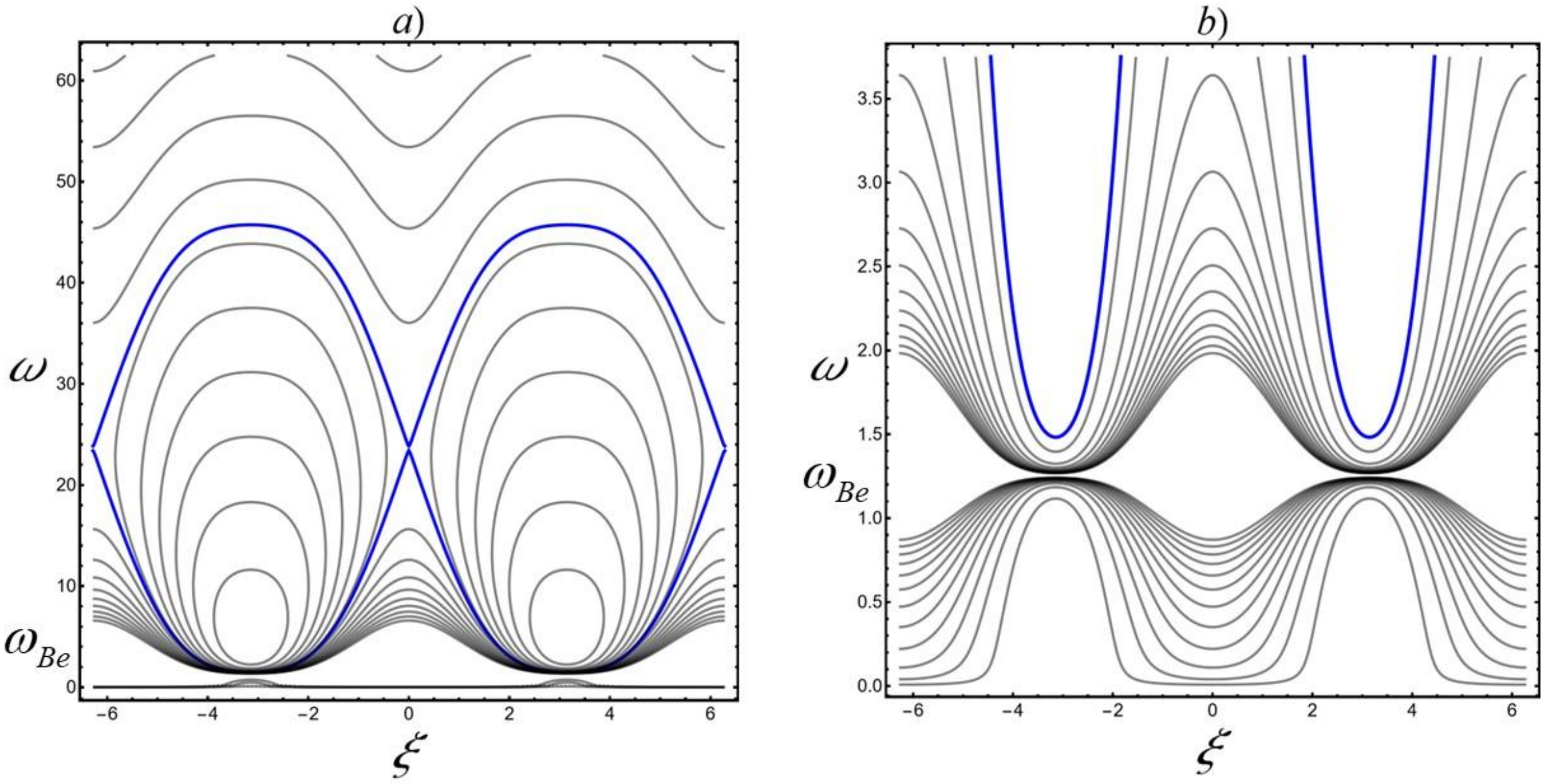

Fig. 5. a) Contours of constant value of the Jacobi integral $\mathcal{H}_R(\xi,\omega)$ given by Eq. (45); b) zoom of the region near $\omega=\omega_{Be}$, for $\beta_w=0.999$, $\omega_{pe}(0)=1$, $\omega_{Be}=1.25$, and $\delta=0.75$. Blue curves stand for the separatrix trajectories.

The stripe near the Larmor frequency where $\omega=\omega_{Be}$ ($\omega_{Be}$ is determined by Eq. (42)) subdivides the phase plane into two subdomains $\omega<\omega_{Be}$ and $\omega>\omega_{Be}$. If the electromagnetic pulse is injected into the photon accelerator with the frequency $\omega\ll\omega_{Be}$, its frequency upshift is limited by the value equal to the Larmor frequency: $\omega_m\le\omega_{Be}$. In the case, the wave packet is injected into the photon accelerator with the frequency $\omega\approx\omega_{Be}$ the maximal frequency upshift is achieved when the packet trajectory is the separatrix (it is a blue curve in the plane $(\xi,\omega)$). Since in the limit $\beta_w\to 1$ the electromagnetic wave frequency is well above that we can find the coordinates of the point of the separatrix branch crossing, assuming smallness of the ratio $\omega_{Be}/\omega$, as

$$\xi_X=2\pi n,\quad \omega_{X,R/L}\approx\frac{\omega_{pe}(\pi n)}{\sqrt{1-\beta_w^2}}\pm\frac{1+\beta_w^2}{2}\omega_{Be} \tag{48}$$

The point, where the group velocity is equal to the phase velocity of the refraction index modulation, (it is $\beta_w c$), corresponds to the critical point on the separatrix $\omega_X$ where two separatrix branches intersect each other. Substituting this expression into the first integral determined by Eq. (45) we find that the constant $h_R$ for upper critical point (where $\omega_{X,2}>\omega_{Be}$) is equal to

$$h_{R,L} \approx \omega_{pe}(0)\sqrt{1-\beta_w^2} \pm \frac{1-\beta_w^2}{2}\,\omega_{Be}\,. \tag{49}$$

Using this and (45 - 46) expressions, we can obtain the maximum value of upshifted frequency, which corresponds to the frequency on the top point of the separatrix on the $(\xi,\omega)$ plane. It is

$$\omega_{\max} \approx \frac{h_{R,L}}{1-\beta_w} = \omega_{pe}(0)\sqrt{\frac{1+\beta_w}{1-\beta_w}} \pm \frac{1+\beta_w}{2}\,\omega_{Be}\,. \tag{50}$$

Minimal frequency is approximately equal to the Larmor frequency, $\omega_{\min} \approx \omega_{Be}$.

In the lower subdomain on the phase plane, where $\omega < \omega_{Be}$, the minimal frequency here is approximately equal to zero as we can also see in Fig. 5 while the maximum frequency equals the Larmor frequency, $\omega_{\max} = \omega_{Be}$. Thus, for the electromagnetic pulse to reach the maximum upshifted frequency, its initial frequency should be above the value of the cutoff frequency determined by Eq. (43),

$$\frac{1}{2}\left(\sqrt{\omega_{Be}^2 + 4\omega_{pe}^2(\xi)} + \omega_{Be}\right), \tag{51}$$

otherwise, it will end up with the frequency that is below the Larmor frequency.

In Fig. 5 we present the phase plane of the Jacobi integral $\mathcal{H}_R(\xi,\omega)$ given by Eq. (45) for a Larmor frequency that is substantially larger than the Langmuir frequency for $\beta_w = 0.999$, $\omega_{pe}(0) = 1$, $\omega_{Be} = 12.5$, and $\delta = 0.75$. For comparison with the case of zero magnetic field, we plot a dotted red separatrix curve for the same parameters, but zero magnetic field. It is seen that the magnetic field effects result in significant change to the upshifted frequency of the R-wave.

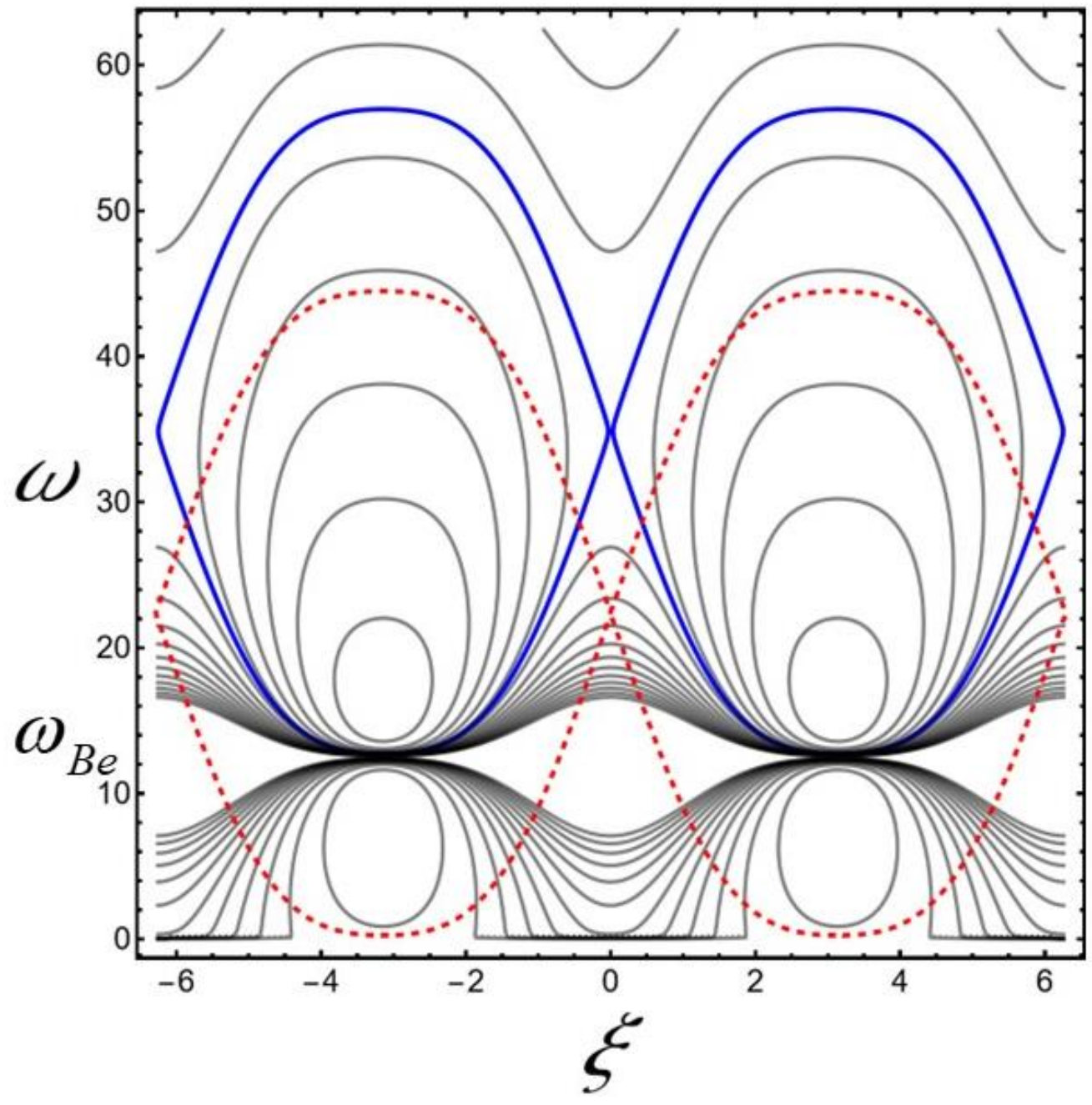

Fig. 6. Contours of constant value of Jacobi integral $\mathcal{H}_R(\xi,\omega)$ given by Eq. (45) for the Larmor frequency substantially larger than the Langmuir frequency for $\beta_w = 0.999$, $\omega_{pe}(0) = 1$, $\omega_{Be} = 12.5$, and $\delta = 0.75$. The red dashed curve shows the separatrix for $\omega_{Be} = 0$.

### *3.2 L-wave*

The phase plot of the Jacobi integral $\mathcal{H}_L(\xi,\omega)$ determined by Eq. (46), for the L-wave is shown in Fig. 7. Here the Langmuir frequency is substantially smaller than the Larmor frequency, $\omega_{pe} \ll \omega_{Be}$. The following parameters are being used: the normalized phase velocity of the refraction index modulations is $\beta_w = 0.999$, the Langmuir frequency at the maximum is $\omega_{pe}(0) = 1$, the Larmor frequency equals to $\omega_{Be} = 2.5$, and the amplitude of the modulations of refraction index is determined by Eq. (12) with the refraction index modulation amplitude determined by $\delta = 0.15$ in frame (a) and $\delta = 0.75$ in frame (b). As in Fig. 1, $\omega_{\min}$ and $\omega_{\max}$ are the minimum and maximum frequency on the separatrix whose critical point is $(0,\omega_X)$.

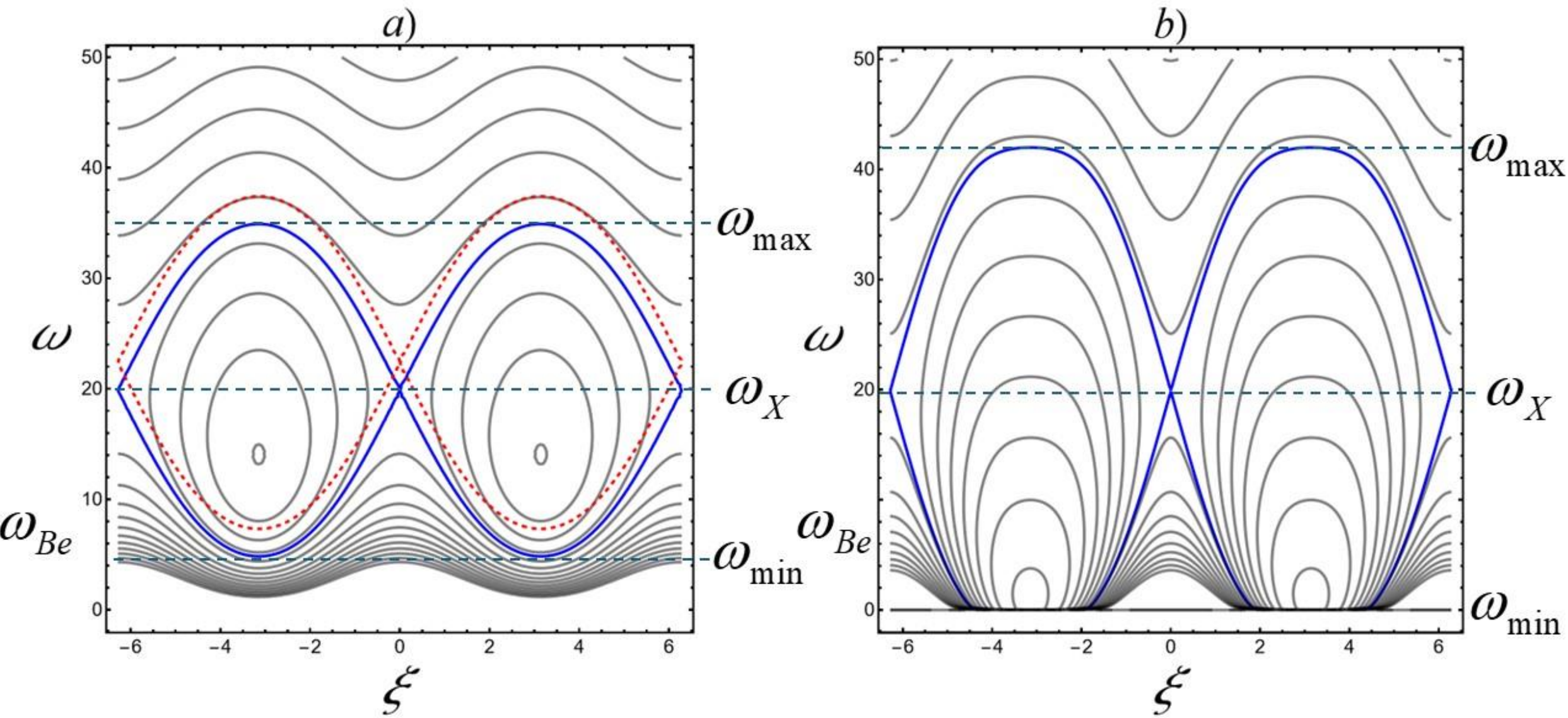

Fig. 7. Contours of constant value of the Jacobi integral $\mathcal{H}_L(\xi,\omega)$ given by Eq. (46) for $\beta_w = 0.999$, $\omega_{pe}(0) = 1$, $\omega_{Be} = 15$, and the refraction index modulation amplitude is determined by $\delta = 0.15$ in frame (a) and $\delta = 0.75$ in frame (b). Here, as in Fig. 1, $\omega_{\min}$, $\omega_{\max}$, and $\omega_X$ are the minimal, maximal frequency on the separatrix (blue curve) whose critical point is $(0,\omega_X)$. Red dashed curve in frame (a) shows the separatrix for $\omega_{Be} = 0$.

In order to find the coordinates of the critical point of the separatrix trajectory on the phase plane $(\xi,\omega)$, where two separatrix branches intersect each other, assuming $\omega_{pe} \ll \omega_{Be}$ we calculate the group velocity of the electromagnetic pulse,

$$v_g(\omega;\xi) \approx c\left(1 - \frac{\omega_{pe}^2(\xi)}{2(\omega+\omega_{Be})^2}\right). \tag{52}$$

At the point $(0,\omega_X)$ the group velocity equals the phase velocity of the refraction index modulations, $c\beta_w$, which gives

$$\omega_X = \frac{\omega_{pe}(0)}{\sqrt{2(1-\beta_w)}} - \omega_{Be}. \tag{53}$$

The value of the Jacobi integral on the separatrix trajectory is

$$h_X = \frac{\omega_{pe}(0)}{\sqrt{2}}\sqrt{1-\beta_w} - \omega_{Be}(1-\beta_w). \tag{54}$$

As a result, we obtain for the maximum and minimum frequencies

$$\omega_{\max/\min} = \frac{\omega_{pe}(0) \pm \sqrt{\omega_{pe}^2(0) - 4\beta_w \omega_{pe}^2(\xi_{\min/\max})}}{2^{3/2}\sqrt{1-\beta_w}} - \omega_{Be} . \tag{55}$$

Here $\xi_{\min}$ is the coordinate where the electromagnetic pulse frequency is at a minimum. In the case of a sufficiently small modulation amplitude, i. e. for $\omega_{pe}^2(0) > 4\beta_w \omega_{pe}^2(\pi/2)$, the minimum and maximum frequency values $\omega_{\max/\min}$ are given by Eq. (57) with $\xi_{\min/\max} = \pi/2k_w$. This case is illustrated by Fig. 7 a).

If $\omega_{pe}^2(0) < 4\beta_w \omega_{pe}^2(\pi/2)$, the minimum frequency equals

$$\omega_{\min} = \frac{\omega_{pe}(0)}{2^{3/2}\sqrt{1-\beta_w}} - \omega_{Be} . \tag{56}$$

For comparison with the case of zero magnetic field, we show in Fig. 7 a) by a dotted red curve a separatrix for the same parameters, but zero magnetic field. It is seen that the magnetic field effects do not lead to significant change of the upshifted frequency of the L-wave.

## 4. Conclusion and Discussions

Theoretical consideration and experimental implementation of the effects of strong magnetic fields in laser accelerators of charged particles and photons open new directions in the development of an understanding of active processes in space active objects and the application of the acquired knowledge in materials science, biology, and fundamental sciences. It was shown that a magnetic field changes the electromagnetic wave interactions with the relativistic plasma wave, it could also lead to an achievable frequency increase within the photon accelerator paradigm.

As noted above, the frequency of the wave packet shifts toward higher frequencies by a factor proportional to the square of the gamma factor corresponding to the velocity of modulation of the refractive index, and due to the Lorentz invariance of the ratio of the electric field of the wave to the frequency of the wave, the amplitude of the electromagnetic wave packet also increases in proportion to the same factor.

The length over which the maximal frequency upshift is reached (it can be called “the photon acceleration length”) is approximately equal to

$$l_{acc} = \frac{2\pi k_w^{-1}}{1-\beta_w} . \tag{57}$$

The effects of a magnetic field lead to quantitative and qualitative changes in the properties of photon acceleration, opening novel opportunities for optimization and increased efficiency. As expected, the most pronounced effect of a magnetic field occurs during the interaction of extraordinary electromagnetic waves with refractive index modulations moving at relativistic phase velocity which can be generated in a plasma imbedded in the static magnetic field in a wake of the driver short pulse laser. Considered above extraordinary electromagnetic waves are the right circularly polarized waves propagating along the static magnetic field in the electron-ion plasma, and the X-wave propagating across the magnetic field. It should be noted that the configuration corresponding to the interaction of high-intensity X-wave with a solid target has shown extremely high efficiency in generating gamma-ray bursts [52]. It has also attracted active attention in connection with the development of the theory of charged particle acceleration and radiation emission from pulsars and magnetars [53].

To demonstrate high electromagnetic wave frequency upshifts ultra-strong magnetic field will require strengths well above the magnetic field achievable with static and pulsed magnets. The self-generated fields in relativistic laser plasma (e, g. see Refs. [54-61] and literature cited therein) are one of candidates. With future advances, coil targets may also be a promising source of strong magnetic fields [62-65].

In laser plasma interactions with near-critical electron density targets (the critical electron density $n_{cr} = m_e \omega^2 / 4\pi e^2$ for one-micron wavelength laser is equal to $\approx 1.2 \times 10^{21} \mathrm{cm}^{-3}$) the Larmor and the Langmuir frequencies are equal to each other for the magnetic field strength of $0.1$ GG. We note that in theoretical/computer simulation publications [62-65] the magnetic field strength is found to reach $10$ GG.

Laboratory measurements of magnetic fields generated during high-intensity laser interactions with dense plasmas [66] showed the generation of magnetic fields with a strength of 0.7 GG localized in a small region with high density near the front surface of solid target. In an extended plasma the magnetic field generated by coil target reached 0.6 KT (i.e. 600 MG) Russell et al. 2025b. As we see, in laser plasma the effects of self-generated magnetic field [66] and magnetic field created with coil-like targets [62-65] can significantly modify the photon accelerator scenario compared with the case of unmagnetized plasma.

## Acknowledgements

The authors thank Tales Augusto Oliveira Gomes, Prokopis Hadjisolomou, Tae Moon Jeong, and Rashid Shaisultanov for fruitful discussions. SSB was supported by U.S. Department of Energy Office of Science Office of High Energy Physics under Contract No. DE-AC02-05CH11231.